\documentclass{elsart}
\usepackage{epsfig,psfig}
\begin{document}

\begin{frontmatter}

\title{Exact Zero-Modes of the Compact QED Dirac Operator}

\author[talla]{Bernd A. Berg,}
\author[talla]{Urs M. Heller,}
\author[wien]{Harald Markum,}
\author[wien]{Rainer Pullirsch,}
\author[wien]{Wolfgang Sakuler}

\address[talla]{Department of Physics\\
 and\\
School of Computational Science and Information Technology,\\
The Florida State University, Tallahassee, FL 32306 }

\address[wien]{Atominstitut der \"Osterreichischen Universit\"aten,
Technische Universit\"at Wien, A-1040 Vienna, Austria}

\begin{abstract}
We calculate the low-lying eigenmodes of the Neuberger overlap-Dirac 
operator for $4d$ compact lattice QED in the quenched approximation. 
In the strong coupling phase we find exact zero-modes, quite similar as in 
non-Abelian lattice QCD. Subsequently we make an attempt to identify
responsible topological excitations of the U(1) lattice gauge theory.
\end{abstract}

\end{frontmatter}

\maketitle

In recent years the spectrum of the Dirac operator in QCD-like theories
has been related to random matrix theory (RMT), see~\cite{Verb00} for 
a review. RMT reproduces the effective, finite-volume partition 
function results of Ref.~\cite{LeSmi92}.
What enters into the RMT description of the low-energy, finite-volume 
scaling behavior are universal symmetry properties of the Dirac operator 
and the topological charge and the value of the chiral condensate 
of the gauge configuration under consideration
\cite{Shur93,Verb94}. Via the Atiyah-Singer index theorem the topological 
charge is mapped on the number of fermionic zero-modes of the Dirac
operator. The symmetry properties of the Dirac operator fall into three 
classes, corresponding to the chiral orthogonal, unitary, and symplectic
ensembles~\cite{Verb94}. This classification is intimately connected to
the chiral symmetry properties of fermions. A good non-perturbative
regularization of QCD should therefore retain those chiral symmetry
properties. Based on the overlap formalism~\cite{NaNe95}, such a
lattice regularization emerged only recently with the massless
Neuberger overlap-Dirac operator~\cite{Neu98} given by
\begin{equation} \label{Neuberger_Dirac}
D = {1\over 2}\, \left[ 1 + \gamma_5\, \epsilon 
\left( H_w(m) \right) \right]\ ,
\end{equation}
where $\gamma_5 H_w(-m)$ is the usual Wilson-Dirac operator on the 
lattice and $\epsilon$ the sign function. The mass $m$ has to be
chosen to be positive and well above the critical mass for Wilson
fermions but below the mass where the doublers become light on the
lattice.

Previously the low-lying spectrum of the operator
(\ref{Neuberger_Dirac}) was calculated for various QCD ensembles
\cite{EHKN99} and agreement with RMT predictions was observed. In this 
paper we expand such calculations to compact, quenched lattice QED 
with the U(1) lattice gauge theory Wilson action
\begin{equation} \label{U1_Wilson}
  S\{U_l\}=\beta\sum_p(1-\cos\theta_p)\ . 
\end{equation}
Here $\beta=1/g^2$, $U_l=U_{x,\mu}=\exp(i\theta_{x,\mu})$,
with $\theta_{x,\mu} \in  (-\pi,+\pi]$ and
$\theta_p=\theta_{x,\mu\nu}=\theta_{x,\mu}+\theta_{x+\hat{\mu},\nu}
-\theta_{x+\hat{\nu},\mu}-\theta_{x,\nu}$ for $\nu\ne\mu$.  For
$\beta<\beta_c\approx 1.01$, this theory is in the confinement phase,
exhibiting a mass gap and monopole excitations~\cite{mono}. This
phase exhibits chiral symmetry breaking and the chiral condensate is,
according to the Banks-Casher relation~\cite{Bank80}, determined by 
the small eigenvalues of the Dirac operator. Properties of the chiral 
phase have been studied numerically in Refs.~\cite{Biel97,Jer,MMP}.
In the strong-coupling limit $\beta\to 0$, chiral symmetry breaking
follows rigorously from infrared bounds~\cite{SaSe91} and it has also
been calculated explicitly~\cite{GaLa91}.  For $\beta>\beta_c$, the
theory is in the chirally symmetric Coulomb phase with a massless 
photon~\cite{Berg84}. There are many interesting questions concerning 
the order of the transition between the two phases and the possibility 
of a non-trivial continuum limit for $\beta\to\beta_c^-$~\cite{Jers96}.

Using the staggered Dirac operator, RMT predictions for U(1) lattice
gauge theory were previously investigated. In both, the strong coupling 
as well as the Coulomb phase, the nearest neighbor spacing distribution of 
the complete eigenvalue spectrum was found to be in agreement 
with the Wigner surmise of the unitary RMT ensemble~\cite{BMP99}, 
indicating quantum chaos. In the strong coupling phase, below the 
Thouless energy, the small eigenvalues were observed to contribute 
to the chiral condensate similarly as for the SU(2) and SU(3) gauge 
groups~\cite{BMP00}, in agreement with the chiral unitary RMT ensemble.

For non-Abelian gauge theories the topological charge is determined
by the instantons~\cite{instantons} of the gauge configuration. 
The topological structure of the Abelian U(1) gauge theory is different.
Therefore,
our first question is whether the overlap-Dirac operator will exhibit
exact zero-modes~\cite{Chiu}. To answer it, we have analyzed configurations
on $L^4$ lattices at $\beta =0.9$ in the confined phase and at 
$\beta =1.1$ in the Coulomb phase. With an overrelaxation/heatbath
algorithm we generated 500 configurations per lattice of linear 
size $L=4$,
$6$ and $8$. After thermalization, the configurations were separated
by 100 sweeps, with each sweep consisting of 3 overrelaxation updates
of each link, followed by one heatbath update. On each configuration
the lowest 12 eigenvalues of the overlap-Dirac 
operator~(\ref{Neuberger_Dirac})
were calculated using the the Ritz functional algorithm~\cite{Ritz}
with the optimal rational approximation of Ref.~\cite{EHN99} to 
$\epsilon \left( H_w(m) \right)$ and $m$ set to 2.3. 

In the confined phase
exact zero-modes of the operator~(\ref{Neuberger_Dirac}) were indeed
found and are compiled in Table~\ref{tab_eigen}. There $n_0^{(\nu)}$ 
denotes the number of configurations on which we found a zero-mode of 
degeneracy $\nu$; $n_0^{(0)}$ is the number of configurations with no 
zero-mode. The total number of zero-modes in all produced configurations 
(of a given lattice size) is
\begin{equation} \label{nzero}
n_0^{\rm tot}\ =\ \sum_{\nu=1}^{\infty} \nu\, n_0^{(\nu)}\ .
\end{equation}
The highest degeneracy observed was $\nu =3$. No zero-modes were found
in the Coulomb phase.

\begin{table}[ht]
\centering
\begin{tabular}{||c|c|c|c|c|c|c||}         \hline
$L$&$n_0^{(0)}$&$n_0^{(1)}$&$n_0^{(2)}$&$n_0^{(3)}$&$n_0^{\rm tot}$&
$ 10^4\, \langle \nu^2\rangle / L^4$ \\ \hline
 4 & 382   & 118   &   0   &   0  &  118 & 9.22 \\ \hline
 6 & 284   & 199   &  17   &   0  &  233 & 4.12  \\ \hline
 8 & 181   & 255   &  61   &   3  &  386 & 2.57  \\ \hline
$Q$& 0.39  & 0.71  & 0.69  &  $-$ &      &                \\ \hline
\end{tabular}
\caption{Exact zero-modes found in 500 configurations per lattice size 
$L$. The column $n_0^{\nu}$, $\nu=0,\dots ,3$, gives the number of 
configurations with $\nu$ zero-modes, $n_0^{\rm tot}$ is the total
number of zero-modes and the last column presents the zero-mode
susceptibility. The last row gives the
$Q$ values of the two-sided Kolmogorov test~\cite{Kolmo} and shows 
consistency of the RMT distributions~(\ref{rho_nu}) with our data.
\label{tab_eigen}}
\end{table}
%
%

For each topological sector $\nu$, chiral RMT predicts the distribution
of the lowest non-zero eigenvalue $\lambda_{\min}$ in terms of the
rescaled variable
\begin{equation} \label{z_def}
z = \Sigma\,V\,\lambda_{\min}\ ,
\end{equation}
where $V$ is the volume of the system and $\Sigma$ the infinite volume
value of the chiral condensate $\langle\overline{\psi}\psi\rangle$
determined up to an overall wave function renormalization. For the U(1) 
gauge group the unitary ensemble applies and the RMT predictions for the
$\nu=0,1,2$ probability densities $\rho^{(\nu)}_{\min}(z)$ 
are~\cite{Fo93,NDW98}
\begin{equation} \label{rho_nu}
\rho^{(\nu)}_{\min}(z)={z\over 2}\, \cases{ e^{-z^2/4}~~{\rm for}~~\nu=0,\cr
I_2(z)\,e^{-z^2/4}~~{\rm for}~~\nu=1,\cr
[I_2(z)^2-I_1(z)I_3(z)]\,e^{-z^2/4}~~{\rm for}~~\nu=2.}
\end{equation}
The chiral condensate is related to the expectation value of 
the smallest eigenvalue. For degeneracy $\nu$ we have
\begin{equation} \label{Sigma}
 \Sigma\ =\ \Sigma^{(\nu)}\ =\ {\langle z^{(\nu)}\rangle \over V\, 
 \langle\lambda^{(\nu)}_{\min}\rangle} \ ,
\end{equation}
where the result is supposed to be independent of $\nu$ and
\begin{equation} \label{average_z}
 \langle z^{(\nu)} \rangle = \int_0^{\infty} dz\, z\, 
 \rho^{(\nu)}_{\min} (z)\ .
\end{equation}
Using analytical and numerical integration the expectation values 
$\langle z^{(\nu)} \rangle$ are easily calculated. One finds
\begin{equation} \label{av_z0_z1}
\langle z^{(0)} \rangle = 1.772453851\,,\ \
\langle z^{(1)} \rangle = 3.107798700
\end{equation}
\begin{equation} \label{av_z2}
{\rm and} ~~~ \langle z^{(2)} \rangle = 4.344018125\ . 
\end{equation}
Using these numbers, the values $\Sigma^{(\nu)}$ follow from Eq.~(\ref{Sigma})
and Table~\ref{tab_Sigma} collects, for the various lattice
sizes and $\nu$'s, the results which we obtain from our data. Within our 
limited statistical accuracy, and allowing for some finite-size 
effects, the different $\Sigma^{(\nu)}$ values are consistent with
one another.

\begin{table}[ht]
\centering
\begin{tabular}{||c|c|c|c|c||}         \hline
$L$& $\Sigma^{(0)}$  & $\Sigma^{(1)}$  & $\Sigma^{(2)}$ & $\Sigma$    \\ \hline
 4 & 0.1264 (36) & 0.1379 (47) & $-$        & 0.1291 (30) \\ \hline
 6 & 0.1342 (43) & 0.1322 (34) & 0.1179 (66)& 0.1328 (28) \\ \hline
 8 & 0.1144 (44) & 0.1290 (36) & 0.1335 (59)& 0.1251 (25) \\ \hline
\end{tabular}
\caption{Estimates of the chiral condensate $\Sigma^{(\nu)}$, where
the subscript $\nu$ denotes the topological sector in which the
estimate is obtained. The last column gives the weighted average
over all topological sectors. \label{tab_Sigma}}
\end{table}

\begin{figure}[-t]
  \begin{center}
    \epsfig{figure=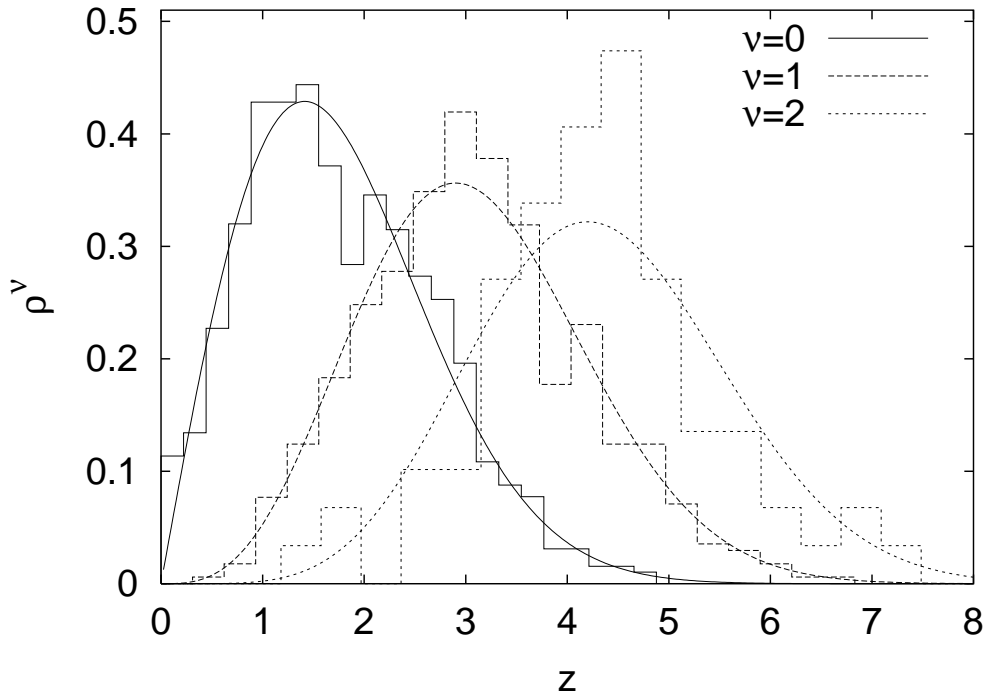,width=\columnwidth}
    \vspace*{-1mm}
    \caption{The exact RMT probability densities~(\ref{rho_nu})
      for the lowest non-zero eigenvalues are compared with
      histograms from our data.}
    \label{fig_pd}
  \end{center}
\end{figure}

Next, we compare the analytical distributions 
$\rho_{\min}^{\nu}$~(\ref{rho_nu}) with our computation.  For $\nu=0$, 
$1$ and $2$, Fig.~\ref{fig_pd} displays the exact RMT probability 
densities $\rho^{(\nu)}_{\min}$ of Eq.~(\ref{rho_nu}) and the 
corresponding histograms from our data. Using the 
definition~(\ref{z_def}) of the $z$-variable and for $\Sigma$ the 
$\Sigma^{(\nu)}$ values
of Table~\ref{tab_Sigma}, all lattice sizes are combined for each
$\nu$ and the comparison is parameter free. Figure~\ref{fig_pd}
shows that the histograms follow the shift of the RMT probability
densities and their general shape. The high peak of the $\nu=2$
histogram is interpreted as a statistical fluctuation and consistent
with the low statistics we have for this case. This claim is made
quantitative by applying the two-sided Kolmogorov test~\cite{Kolmo} 
to the discrepancy between the RMT and our empirical distributions, 
which leads to the satisfactory $Q$ values in the last row of 
Table~\ref{tab_eigen}. 
The topological structure of the U(1) gauge theory is not as 
extensively studied as the non-Abelian case. Nevertheless, 
with torons~\cite{torons}, monopole solutions~\cite{mono} and Dirac
sheets~\cite{GJJ85} a number of topological objects are known for
the U(1) gauge theory and their possible relationship to the zero-modes
of the overlap-Dirac operator is discussed in the remainder
of this paper.

{\it Torons}. For a fixed link $(x,\hat{\mu})$ we average the angle
$\theta_{x,\mu}$ over the perpendicular 3d space to obtain
$\overline{\theta}^{\,\mu}_i$,
$i=0,\dots,L-1$.
We define the toron charge for direction $\hat{\mu}$ as the sum,
using periodic boundary conditions,
\begin{equation} \label{q_tor^mu}
 q^{\mu}_{\rm tor}\ =\ {1\over 2\pi} \sum_{i=1}^L
\overline{\theta}_{i,i-1}^{\,\mu}\ ,
\end{equation}
where $\overline{\theta}_{i,i-1}^{\,\mu}$ is the shorter distance between
$\overline{\theta}_{i}^{\,\mu}$ and $\overline{\theta}_{i-1}^{\,\mu}$ on
the circle, yielding an integer quantity.
Table~\ref{tab_torons} gives the distributions
of the total toron charges
\begin{equation} \label{q_tor}
 q_{\rm tor}\ =\ \sum_{\mu=1}^4 q^{\mu}_{\rm tor}
\end{equation}
for all our configurations at $\beta=0.9$ in the confined phase. No non-zero
toron charges are found at $\beta=1.1$ in the Coulomb phase.

\begin{table}[ht]
\centering
\begin{tabular}{||c|c|c|c|c|c|c|c|c|c|c|c||}                  \hline
$L\ \backslash\ q$& -5& -4& -3& -2& -1& 0 & +1& +2& +3& +4& +5\\ \hline
 4~~~~~           &  0&  1&  7& 37&111&193&117& 26&  8&  0&  0\\ \hline
 6~~~~~           &  0&  1& 12& 56&110&142&107& 51& 21&  0&  0\\ \hline
 8~~~~~           &  0&  5& 32& 49& 88&134&113& 51& 22&  5&  1\\ \hline
\end{tabular}
\caption{Distribution of the total toron charge~(\ref{q_tor}). 
\label{tab_torons}} \end{table}

At $\beta =0.9$ we checked for correlations between toron charges
$q^{\mu}_{\rm tor}$ and $q_{\rm tor}$ with the degeneracy of our
exact zero-modes and found none within the limitations of our
statistical errors. Table~\ref{tab_md} includes the average number
of torons  versus $\nu$ (i.e. the average of the absolute values
of the charges listed in Table~\ref{tab_torons}).
From $L=4$ to $6$ there is an increase with
lattice size, but no statistically significant dependence on $\nu$.

{\it Monopoles and Dirac sheets}.
The U(1) plaquette angles $\theta_{x,\mu\nu}$ are decomposed into the
``physical'' electromagnetic flux through the plaquette $\bar \theta_{x,\mu\nu}$
and a number $m_{x,\mu\nu}$ of Dirac strings passing through the
plaquette
\begin{equation} \label{Dirac_string_def}
 \theta_{x,\mu\nu} = \bar \theta_{x,\mu\nu} + 2\pi\,m_{x,\mu\nu}\ ,
\end{equation}
where $\bar \theta_{x,\mu\nu}\in (-\pi,+\pi]$. We shall call plaquettes
with $m_{x,\mu\nu} \ne 0$ {\it Dirac plaquettes}. The monopole charges in
elementary $3d$ cubes are defined as the net number of Dirac strings
entering or exiting these cubes. The worldlines of these monopoles on
the dual lattice are closed, either within the lattice volume or by the
periodic boundary conditions. The dual integer valued plaquettes
\begin{equation} \label{monopole_def}
m^*_{x,\mu\nu} = \frac{1}{2} \varepsilon_{\mu\nu\rho\sigma} m_{x,\rho\sigma}
\end{equation}
form Dirac sheets bounded by the worldlines of monopole-antimonopole
pairs. Due to the periodic boundary conditions they can also be closed
without the presence of any monopoles or antimonopoles. In the remainder
we shall refer to these as {\it Dirac sheets}. A Dirac sheet in the
x-y plane of the $4d$ lattice must contain at least $L_z L_t$ Dirac
plaquettes. Of course, monopole loops can form ``holes'' in these Dirac
sheets. So, in practice we defined the number of Dirac sheets in the
x-y plane as the closest integer to $N_{1,2} / (L_z L_t)$ where $N_{1,2}$
is the total number of Dirac plaquettes in the x-y planes.

\begin{table}[ht]
\centering
\begin{tabular}{||c|c|c|c|c||}                     \hline
$L$ &$\nu$ & Torons    &  Monopoles    &   Dirac Sheets  \\ \hline
 4  & 0    & 0.81 (04) &  198.6 (1.2)  &   0.099 (16)    \\ \hline
 4  & 1    & 0.81 (06) &  198.9 (2.0)  &   0.136 (32)    \\ \hline
 6  & 0    & 1.06 (05) & 1007.5 (2.9)  &   0.106 (19)    \\ \hline
 6  & 1    & 1.08 (07) & 1007.7 (3.3)  &   0.101 (22)    \\ \hline
 6  & 2    & 1.18 (15) & 1015    (16)  &   0.112 (81)    \\ \hline
 8  & 0    & 1.17 (07) & 3183.9 (5.7)  &   0.100 (21)    \\ \hline
 8  & 1    & 1.19 (07) & 3195.8 (6.2)  &   0.100 (20)    \\ \hline
 8  & 2    & 1.46 (14) & 3197   (12)   &   0.119 (44)    \\ \hline
 8  & 3    & 1.33 (88) & 3221   (18)   &   none          \\ \hline
\end{tabular}
\caption{Average numbers of torons, monopoles and Dirac sheets per
configuration with fixed zero-mode degeneracy $\nu$. \label{tab_md}}
\end{table}

In the confined phase we find a large number of monopoles whereas the
typical Dirac sheet numbers are $0$ and $\pm 1$ (only on the $8^4$
lattices we found at $\nu=0$ a single configuration with two Dirac
sheets). The average numbers of monopoles and Dirac sheets do not
correlate with the topological charge $\nu$ (i.e. our zero-mode
degeneracy) as the numbers of Table~\ref{tab_md} show. One might
observe a slight increase of the monopole number with $\nu$ which
has no statistical significance.
The average number of Dirac sheets shows no lattice size dependence
within the limits of our statistics. In the Coulomb phase we find
few monopoles and no Dirac sheets. 

All these topological
configurations have in common with the zero-mode degeneracy that
they are turned on at the phase transition from order (Coulomb)
to disorder (confined). However, in the confined phase  we found
no detailed correlation between any of the topological phenomena
and the zero-mode degeneracy of the overlap-Dirac operator. This
might be related to the fact that the zero-mode susceptibility,
depicted in the last column of table~\ref{tab_eigen}, decreases and 
may approach zero for $L\to\infty$.

In contrast to these statistical findings there is an interesting
observation that ordered Dirac sheet configurations give rise to
zero-modes of the overlap-Dirac operator, with the number of zero-modes
being equal to the number of Dirac sheets~\cite{NaNe95,Chiu}.
One might understand this from the fact that a Dirac sheet is a
$2d$ gauge configuration that contains unit topological charge in
the $2d$ sense, kept constant in the two orthogonal directions.
One might even verify this coincidence~\cite{NaNe95} by cooling
or smoothing of equilibrium gauge fields~\cite{TFMS96}.

\begin{figure*}[t]
\centerline{\hspace*{15mm}\psfig{figure=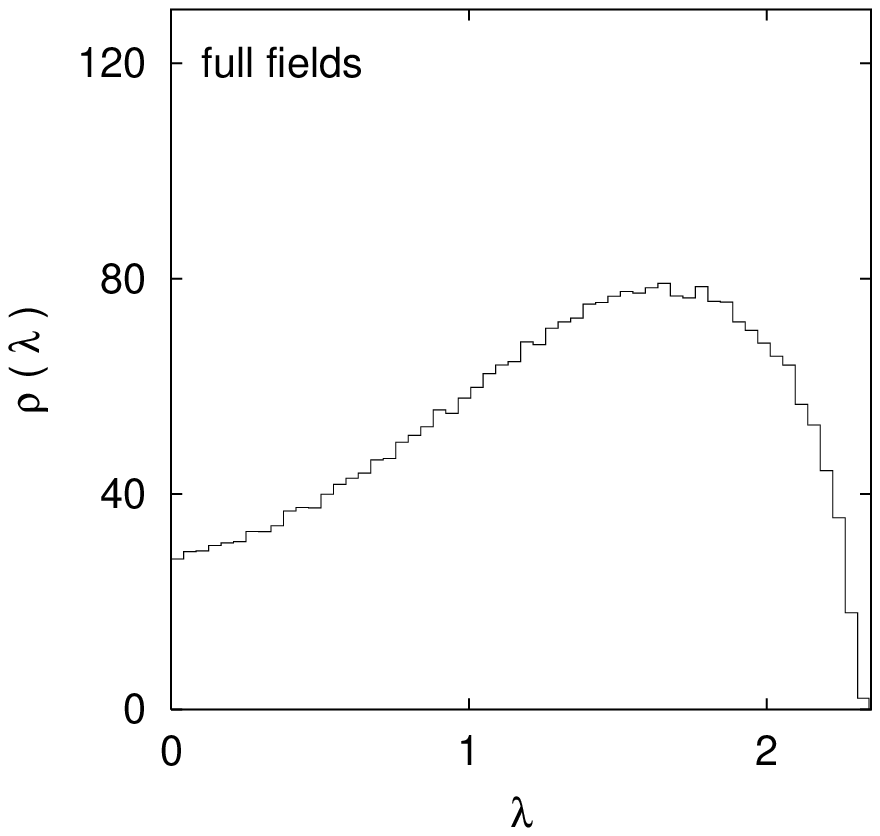,width=6cm,height=4cm}\hspace*{-15mm}
            \psfig{figure=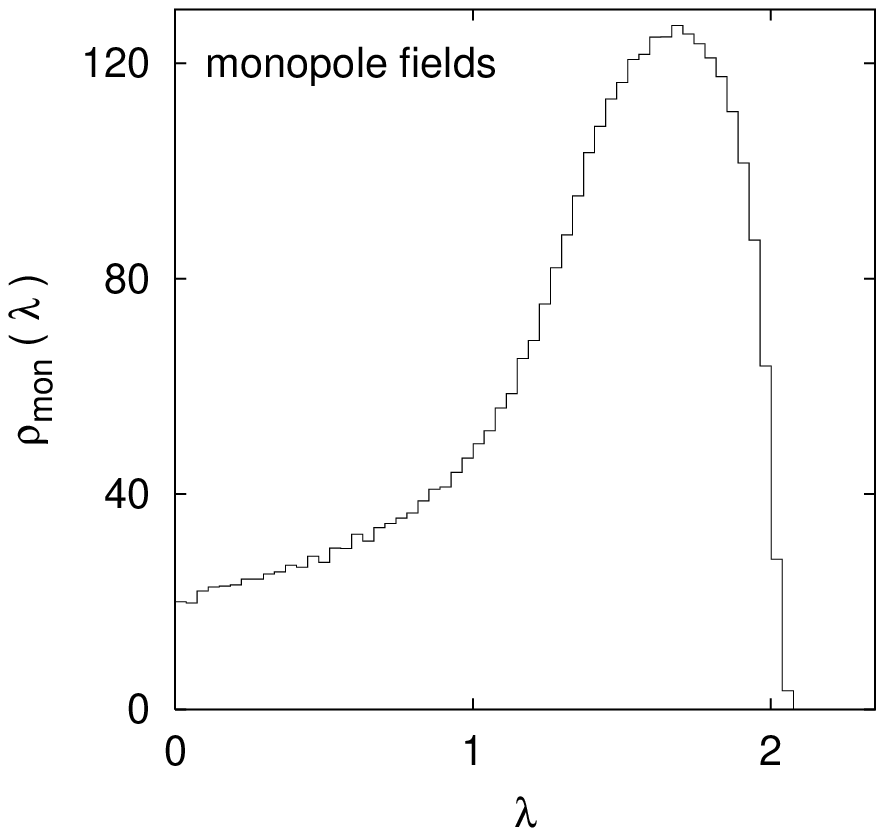,width=6cm,height=4cm}\hspace*{-15mm}
            \psfig{figure=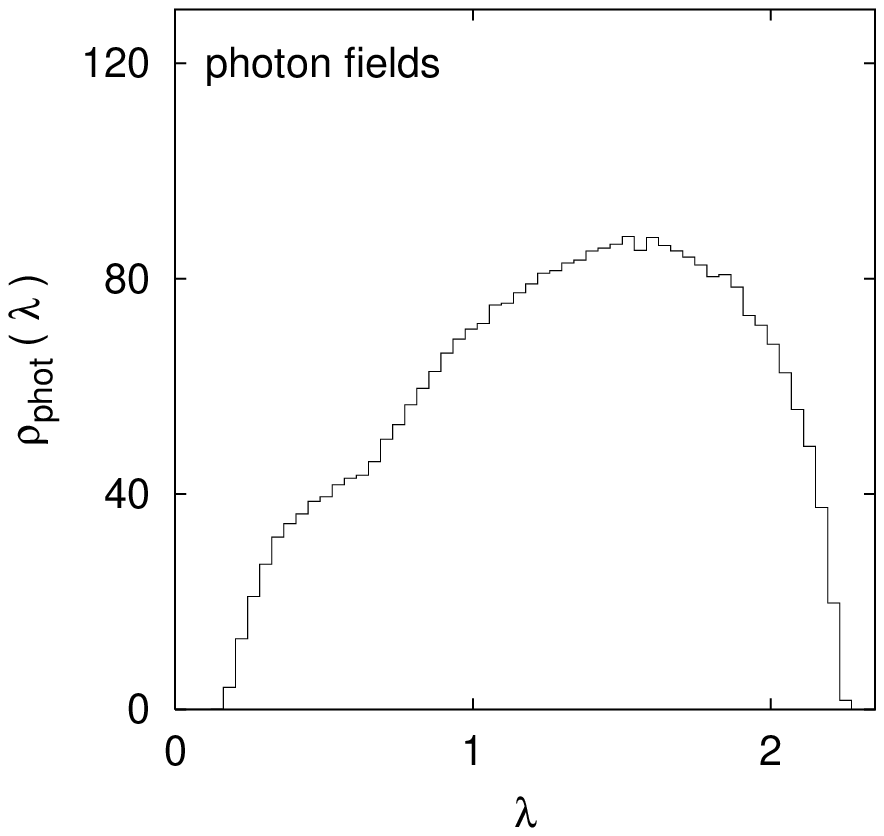,width=6cm,height=4cm\hspace*{-0mm}}}
    \vspace*{-1mm}
    \caption{Decomposition of the spectrum $\rho(\lambda)$ (normalized to the
             number of eigenvalues) of the staggered Dirac operator into a
             monopole part $\rho_{\rm mon}(\lambda)$ and a photon part
             $\rho_{\rm phot}(\lambda)$, averaged over 500 configurations
             on a $4^4$ lattice.}
    \label{staggered}
\end{figure*}

To investigate these correlations further we have, following
Refs.~\cite{StWe92,Suzu96,Biel97}, factorized our gauge configurations
into monopole and photon fields in the following way
\begin{equation} \label{monoplaq}
\theta^{\rm mon}_{x, \mu} = - 2 \pi\, \sum_{x'} G_{x,x'} \,
\partial_{\nu}' \, m_{x', \nu\mu}
\end{equation}
\begin{equation} \label{photplaq}
\theta^{\rm phot}_{x, \mu} = - \, \sum_{x'} G_{x,x'} \,
\partial_{\nu}' \, \bar\theta_{x', \nu\mu} \ .
\end{equation}
Here $\partial'$ acts on $x'$, the quantities
$m_{x,\mu\nu}$ and $\bar\theta_{x, \mu\nu}$ are defined in
Eq.~(\ref{Dirac_string_def}) and 
\begin{equation}
G_{x,x'} = \sum_{p_{\mu} (p\ne0)} \, \frac{1}{\sum_{\nu} 4 \sin^{2}
(\frac{p_{\nu}a}{2})}  \, \exp\left[ \sum_{\mu} i \, p_{\mu} (x_{\mu} - x'_{\mu}) \right]
\end{equation}
is the lattice Coulomb 
propagator. One can show that $\theta_{x,\mu}=\theta^{\rm mon}_{x,\mu}
+\theta^{\rm phot}_{x,\mu}$ is up to a gauge transformation identical
with the original $\theta_{x,\mu}$ defined by $U_{x,\mu}=
\exp(i \theta_{x,\mu})$.

This is a different way to probe the topological content 
of a gauge theory and we computed so far the total density of 
eigenvalues, with $\int \rho(\lambda) d \lambda = V$.
We found that the zero-modes lie solely in the monopole 
part of the gauge field and are completely absent in the photonic 
field.  Using periodic boundary conditions in space and anti-periodic 
boundary conditions in time~\cite{Biel97}, this was seen both for the 
overlap-Dirac operator and for the quasi-zero-modes of the staggered 
Dirac operator, see Fig.~\ref{staggered}. 
At the moment we are accumulating statistics to perform a decomposition
into different topological sectors and to obtain an analogous analysis
as we did with the original U(1) field above. It is of further interest
to study space-time correlations between different topological objects.
We pose the question of the existence of local correlations between the
topological charge density and the monopole density. We plan to 
calculate spatial correlations with the wave functions related to fixed 
zero-mode number $\nu$ and hope to understand the confining mechanism 
of compact QED.

\hfill\break

{\it Acknowledgments:}
This work was supported in part by the US Department of Energy under
contract DE-FG02-97ER41022 and by the Fonds zur F\"orderung der 
wissenschaftlichen Forschung under project P14435-TPH.

\end{document}